\documentstyle[11pt,amssymb,epsf]{article}

\def\nn{\nonumber}

\let\bm=\bibitem

\def\be{\begin{equation}}
\def\ee{\end{equation}}
\def\ba{\begin{array}}
\def\ea{\end{array}}

\def\fft#1#2{\frac{#1}{#2}}
\def\del{\partial}

\def\sst#1{{\scriptscriptstyle #1}}

\def\td{\tilde}
\def\wtd{\widetilde}
\def\ie{\rm i.e.\ }
\def\dalemb#1#2{{\vbox{\hrule height .#2pt
        \hbox{\vrule width.#2pt height#1pt \kern#1pt
                \vrule width.#2pt}
        \hrule height.#2pt}}}

\newcommand{\hoch}[1]{$\, ^{#1}$}
\newcommand{\bea}{\begin{eqnarray}}
\newcommand{\eea}{\end{eqnarray}}

\def\0{{\sst{(0)}}}
\def\1{{\sst{(1)}}}
\def\2{{\sst{(2)}}}
\def\3{{\sst{(3)}}}
\def\4{{\sst{(4)}}}
\def\5{{\sst{(5)}}}
\def\6{{\sst{(6)}}}
\def\7{{\sst{(7)}}}
\def\8{{\sst{(8)}}}

\def\im{{{\rm i}}}

\begin{document}
\begin{flushright}
MIFP-06-10 \\
{\bf hep-th/0604125}\\
April\  2006
\end{flushright}

\begin{center}

{\Large {\bf General Kerr-NUT-AdS Metrics in All Dimensions}}

\vspace{20pt}

W. Chen, H. L\"u and C.N. Pope

\vspace{20pt}

{\hoch{\dagger}\it George P. \&  Cynthia W. Mitchell Institute
for Fundamental Physics,\\
Texas A\&M University, College Station, TX 77843-4242, USA}

\vspace{40pt}

\underline{ABSTRACT}
\end{center}

   The Kerr-AdS metric in dimension $D$ has cohomogeneity $[D/2]$; the
metric components depend on the radial coordinate $r$ and $[D/2]$ latitude
variables $\mu_i$ that are subject to the constraint $\sum_i \mu_i^2=1$.
We find a coordinate reparameterisation in which the $\mu_i$ variables
are replaced by $[D/2]-1$ unconstrained coordinates $y_\alpha$, and having
the remarkable property that the Kerr-AdS metric becomes diagonal in the
coordinate differentials $dy_\alpha$.  The coordinates $r$ and $y_\alpha$
now appear in a very symmetrical way in the metric, leading to an immediate
generalisation in which we can introduce $[D/2]-1$ NUT parameters.
We find that $(D-5)/2$ are non-trivial in odd dimensions, whilst
$(D-2)/2$ are non-trivial in even dimensions.
This gives the most general Kerr-NUT-AdS metric in $D$ dimensions.  We 
find that in all dimensions $D\ge4$ there exist discrete 
symmetries that involve inverting a rotation parameter through the AdS
radius.  These symmetries imply that Kerr-NUT-AdS metrics with over-rotating
parameters are equivalent to under-rotating metrics.  
We also consider the BPS limit of the Kerr-NUT-AdS
metrics, and thereby obtain, in odd dimensions and after Euclideanisation, 
new families of Einstein-Sasaki metrics.

{\vfill\leftline{}\vfill \vskip 10pt \footnoterule {\footnotesize
Research supported in part by DOE grant
DE-FG03-95ER40917.
}


\newpage
\tableofcontents
\addtocontents{toc}{\protect\setcounter{tocdepth}{2}}

\section{Introduction}

    With the discovery of higher-dimensional supergravities, string theory
and M-theory, it has been appreciated that higher-dimensional solutions
of the Einstein and supergravity equations have an important r\^ole to
play.  With this realisation, it becomes of considerable interest to
carry out investigations in higher dimensions, generalising the earlier
very extensive four-dimensional studies, of the most general classes
of solutions subject to certain symmetry assumptions.  In this regard,
black hole metrics and their generalisations are amongst the most
important solutions worthy of investigation.

   In four dimensions the early discovery of Schwarzschild solution was
quickly followed by the extension to the Reissner-Nordstr\"om solution,
and, considerably later, by the discovery of generalisations with
rotation, a cosmological constant, a NUT parameter, and an acceleration
parameter.  The general solution containing all these parameters was obtained
in \cite{plebdemi}.  Less is known in higher dimensions, and it was not until
1986 \cite{myeper} that the generalisation of the Kerr solution to dimensions
$D\ge 5$ was obtained.  A feature of these higher-dimensional black
holes is that there are $[(D-1)/2]$ independent rotation parameters,
corresponding to independent rotations in the $[(D-1)/2]$ orthogonal
spatial 2-planes.  The metrics in general have cohomogeneity $[D/2]$ in $D$
dimensions, with lower cohomogeneities arising in specialisations where
some of the rotation parameters are set equal.
The generalisation to include a cosmological
constant was obtained in five dimensions in \cite{hawhuntay}, and
in arbitrary dimension in \cite{gilupapo1,gilupapo2}.

   It is of interest to generalise these higher-dimensional Kerr-AdS
metrics further, by including also NUT charge.  Recent progress in this
direction was made in \cite{chlupo1}, where it was shown that one can
introduce a NUT charge parameter in all the
Kerr-AdS metrics in $D\ge 6$ if they are first specialised, by equating
rotation parameters appropriately, to have cohomogeneity 2.  The case
of $D=4$ had, of course, been obtained long ago, and the case $D=5$
turns out to be rather degenerate, in that the NUT parameter is
trivial and can be removed by a redefinition of the other parameters
and the coordinates. (Cohomogeneity-one pure multi-nut solutions
in higher dimensions were obtained in \cite{mann}.)

   The purpose of the present paper is to present new NUT generalisations
of the Kerr-AdS metrics which are, we believe, of the most general
possible type.  We find that in $D$ dimensions the general Kerr-AdS
metric (with all rotation parameters allowed to be unequal) can be
extended by the inclusion of $(D-5)/2$ independent NUT parameters when
$D$ is odd, and $(D-2)/2$ when $D$ is even.
We arrived at these solutions by first rewriting the Kerr-AdS
metrics using a set of coordinate variables that make the introduction
of the NUT parameters a very natural generalisation of the usual mass
parameter.

   The choice of coordinates in four dimensions that leads to the
natural inclusion of a NUT parameter in the Kerr-AdS solution is rather
well known.  In the standard description of the Kerr-AdS solution
one has angular coordinates $(\theta,\phi)$ parameterising the
2-sphere spatial sections at constant radius $r$.  If one defines
$y=a \cos\theta$, where $a$ is the rotation parameter, and makes
appropriate linear redefinitions of the time and azimuthal coordinate $\phi$,
the metric can be written as
\be
ds^2_4= -\fft{\Delta_r}{r^2+u^2}\, (d\tau + y^2 d\psi)^2 +
   \fft{\Delta_y}{r^2+u^2}\, (d\tau - r^2 d\psi)^2 +
  \fft{(r^2+y^2)\, dr^2}{\Delta_r} + \fft{(r^2+y^2)\, dy^2}{\Delta_y}\,,
\ee
where
\be
\Delta_r = (r^2+a^2)(1+g^2 r^2) -2M r\,,\qquad
\Delta_y= (a^2-y^2)(1-g^2 y^2)\,.
\ee
This Kerr-AdS solution, satisfying $R_{\mu\nu}= -3 g^2\, g_{\mu\nu}$, is
generalised to include the NUT parameter $L$ by replacing $\Delta_y$ by
\be
\Delta_y= (a^2-y^2)(1-g^2 y^2) + 2 L y\,.
\ee

   An important feature of this parameterisation, which makes the inclusion of
the NUT parameter very natural, is that the radial variable $r$ and the
``latitude'' variable $y$ are placed on a very symmetrical footing.  The
NUT generalisations of the higher-dimensional Kerr-AdS metrics that were
obtained in \cite{chlupo1} worked in a very similar way.  An essential
part of the construction was that the rotation parameters had to be
specialised in such a way that the cohomogeneity was reduced to 2, and
so again a latitude-type coordinate $y$ could be introduced in such a
way that it, and the radial variable $r$, appeared in a very symmetrical
way.  The metric functions depended on $r$ and $y$, with the
$(D-2)$-dimensional hypersurfaces at constant $r$ and $y$ being homogeneous.

   The key to finding the NUT generalisations that we obtain in the present
paper is to make a suitable reparameterisation of the multiple ``latitude''
coordinates that arise in the higher-dimensional Kerr-AdS metrics.  In
$D$ dimensions one has the time and radial variables $(t,r)$,
$[(D-1)/2]$ azimuthal angles $\phi_i$ and
$[D/2]$ latitude, or direction cosine, coordinates $\mu_i$, which are
subject to the constraint
\be
\sum_{i=1}^{[D/2]} \mu_i^2 =1\,.\label{mucon}
\ee
The spatial sections at constant radius $r$ have the geometry of deformed
$(D-2)$-spheres.  The unit $S^{D-2}$ metric is given by
\be
d\Omega^2 = \sum_{i=1}^{[D/2]} d\mu_i^2 +
 \sum_{i=1}^{[(D-1)/2]} \mu_i^2 d\phi_i^2 \label{sphere}
\ee
in these variables.  Associated with each azimuthal angle $\phi_i$ is
a rotation parameter $a_i$.

   We find that the appropriate reparameterisation of the $\mu_i$
coordinates is as follows.  Taking $D=2n+1$
in the odd-dimensional case, and $D=2n$ in the even-dimensional case,
we parameterise the $n$ coordinates $\mu_i$ as
\be
\mu_i^2 = \fft{\prod_{\alpha=1}^{n-1} (a_i^2 - y_\alpha^2)}{
     {\prod'}_{k=1}^{n} (a_i^2-a_k^2)}\,,\label{muy}
\ee
where the prime on $\prod'$ indicates that the term that vanishes
(\ie when $k=i$) is omitted from the product.\footnote{Transformations
of this type were first considered by Jacobi, in the context of constrained
dynamical systems \cite{jacobi}.}  Note that this parameterisation
using just $(n-1)$ coordinates $y_\alpha$ explicitly solves the constraint
(\ref{mucon}).  It also has the striking property that it diagonalises
the metric (\ref{sphere}) on the unit sphere, expressed in terms of the
unconstrained latitude variables $y_\alpha$:
\be
d\Omega^2 = \sum_{\alpha=1}^{n-1} g_\alpha\, dy_\alpha^2 +
 \sum_{i=1}^{[(D-1)/2]} \mu_i^2\, d\phi_i^2\,,\label{sphere2}
\ee
where the notation $\prod'$ universally indicates that the vanishing factor
is to be omitted from the product, $\mu_i^2$ is given by(\ref{muy}), and
\be
g_\alpha= - \fft{y_\alpha^2\,
{\prod'}_{\beta=1}^{n-1}
 (y_\alpha^2- y_\beta^2)}{\prod_{k=1}^n
    (a_k^2 - y_\alpha^2)}\,.
\ee
Note that the parameterisation
(\ref{muy}) solves the constraint (\ref{mucon}), and diagonalises the
metric as in (\ref{sphere2}), for arbitrary choices of unequal constants
$a_i^2$.

    When we utilise (\ref{muy}) in the
next section, we shall take the constants $a_i$ to be the rotation
parameters of the Kerr-AdS black holes.  In the case of even dimensions
$D=2n$, there are only $(n-1)$ rotation parameters, and so $a_n$ is taken
to be zero.  We shall see that with this choice of the parameters in
the Jacobi
transformations (\ref{muy}), the Kerr-AdS metrics obtained in
\cite{gilupapo1,gilupapo2}, which are non-diagonal in the latitude
coordinate differentials $d\mu_i$, remarkably become diagonal with respect
to the unconstrained coordinate differentials $dy_\alpha$.  Furthermore,
we shall see that after writing the Kerr-AdS metrics in terms
of the coordinates $(t,r,y_\alpha,\phi_i)$, the radial
variable $r$ and the latitude variables $y_\alpha$ enter the metrics in
a very symmetrical fashion, such that
the generalisation to include a set of $(n-1)$ NUT parameters becomes
very natural.  We have explicitly verified, with the aid of a computer,
that in all dimensions $D\le15$ these generalisations of the Kerr-AdS
metrics satisfy the Einstein equations.  Since there are no special
features peculiar to these dimensions, we can with confidence
expect that the generalisations satisfy the Einstein equations in all
dimensions.

   After presenting the general Kerr-NUT-AdS metrics in section 2, we
then consider, in section 3, some simpler expressions for
the Kerr-NUT-AdS metrics.  It turns out that the symmetrical appearance of
the radial and latitude variables is further enhanced if one performs
a ``Wick rotation'' of the radial coordinate $r$, and defines variables
$x_\mu$ with $x_\alpha=y_\alpha$, $x_n= \im\, r$.  This leads to a
form for the metric in which all the coordinates $x_\mu$ enter on an
exactly parallel footing.  In a further simplification of the 
expressions for the metrics, we find that by defining appropriate 
linear combinations of the time and azimuthal coordinates, the 
Kerr-NUT-AdS metrics can be cast in a form that provides a natural
generalisation of the four-dimensional metrics described in \cite{pleb}.
We also discuss certain scaling symmetries and discrete symmetries of the
Kerr-NUT-AdS metrics.  The scaling symmetries imply that there are 
$(n-2)$ non-trivial NUT parameters in odd dimensions $D=2n+1$, and
$(n-1)$ non-trivial NUT parameters in even dimensions $D=2n$.  The
discrete symmetries imply that metrics with over-rotation, \ie
where one or more rotation parameters exceeds the AdS radius, are
equivalent to metrics with under-rotation.

   In section 4, we focus on the particular cases of dimensions $D=6$ and
$D=7$, since these are the lowest dimensions where our new results
extend beyond those known previously.  In section 5 we study the
supersymmetric, or BPS, limits of the new metrics in odd and even 
dimensions.  After
performing a Euclideanisation, the odd-dimensional solutions give rise to 
new examples of Einstein-Sasaki metrics in $D\ge 7$.  By writing
these as circle fibrations over an Einstein-K\"ahler base, we thereby
obtain new classes of Einstein-K\"ahler metrics in all even dimensions
$D\ge 6$.  The paper ends with conclusions in section 6. 

\section{The General Kerr-NUT-AdS Solutions}

    In this section, we shall present our general results for the
Kerr-NUT-AdS metrics in $D$ dimensions.  These ostensibly have a total of
$(D-1)$ independent parameters, comprising the mass
$M$, the $[(D-1)/2]$ rotation parameters $a_i$, and $[(D-2)/2]$
NUT parameters $L_\alpha$.  As we shall discuss later, in odd 
dimensions there is a symmetry that allows one to eliminate one of 
the parameters, and
so in odd dimensions there are actually in total 
$(D-2)$ non-trivial parameters in the
solutions we obtain.

    The first step is to rewrite the Kerr-AdS metrics, which were obtained
in \cite{gilupapo1,gilupapo2}, in terms of the new coordinate parameterisation
introduced in (\ref{muy}).   It is advantageous to separate the discussion
into two cases, depending upon whether $D$ is odd or even.

\subsection{The odd-dimensional case: $D=2n+1$}

As a preliminary, we make the following
definitions:
\bea
U&=& \prod_{\alpha=1}^{n-1} (r^2+ y_\alpha^2)\,,\qquad
U_\alpha= -(r^2+ y_\alpha^2){{\prod}'}_{\beta=1}^{\, n-1}
(y_\beta^2-y_\alpha^2)
\,,\qquad 1\le \alpha \le n-1\,,\nn\\
W &=&\prod_{\alpha=1}^{n-1} (1-g^2 y_\alpha^2)\,,\qquad
 \gamma_i = \prod_{\alpha=1}^{n-1} (a_i^2 - y_\alpha^2)\,,\qquad
1\le i\le n\,,\nn\\
X &=& \fft{1+g^2 r^2}{r^2}\, \prod_{k=1}^n (r^2 + a_k^2) - 2M\,,\nn\\
X_\alpha &=& \fft{1-g^2 y_\alpha^2}{y_\alpha^2}\, \prod_{k=1}^n
    (a_k^2 - y_\alpha^2) + 2 L_\alpha\,\qquad
1\le \alpha\le n-1\,.\label{functions}
\eea
We have actually already included the new NUT parameters $L_\alpha$ here;
they appear just in the definitions of the functions $X_\alpha$.  Note that
again the notation $\prod'$ indicates that the term in the full product that
vanishes is to be omitted.

    Using these functions, we find that the Kerr-NUT-AdS metrics in
$D=2n+1$ dimensions are given by
\bea
ds^2 &=&  \fft{U}{X}\, dr^2
+\sum_{\alpha=1}^{n-1} \fft{U_\alpha}{X_\alpha}\, dy_\alpha^2
- \fft{X}{U}\,
\Big[Wd \td t - \sum_{i=1}^n a_i^2 \gamma_i d\td\phi_i\Big]^2\nn\\
&&
+ \sum_{\alpha=1}^{n-1} \fft{X_\alpha}{U_\alpha}\,
\Big[ \fft{(1+g^2 r^2) W}{1- g^2 y_\alpha^2}\, d\td t -
  \sum_{i=1}^n \fft{a_i^2 (r^2 +a_i^2)\, \gamma_i}{a_i^2- y_\alpha^2}\,
  d\td\phi_i \Big]^2 \nn\\
&&
+ \fft{\prod_{k=1}^n a_k^2}{r^2\, \prod_{\alpha=1}^{n-1} y_\alpha^2}\,
\Big[(1+g^2 r^2) W\, d\td t - \sum_{i=1}^n (r^2+a_i^2)\gamma_i\, d\td\phi_i
    \Big]^2\,.\label{oddmetric}
\eea
With the parameters $L_\alpha$ set to zero, the metrics are just a rewriting
of the Kerr-AdS metrics obtained in \cite{gilupapo1,gilupapo2}, using the
new coordinates $y_\alpha$ defined by (\ref{muy}).\footnote{Note
that the metric signature is just the usual $(-++\cdots +)$, for
the appropriate choices of the $y_\alpha$ coordinate intervals that
correspond to the standard Kerr-AdS black hole solution.}  They are
written here in an asymptotically-static frame.  We have also
rescaled the time and azimuthal coordinates in order to simplify the
expression.  They are related to the original asymptotically static
coordinates $(t,\phi_i)$ by
\be
t= \td t\, \prod_{i=1}^n \Xi_i\,,\qquad \phi_i= a_i\, \Xi_i\, \td\phi_i\,
       {{\prod}'}_{k=1}^{\, n} (a_i^2-a_k^2)\,,\label{ttp}
\ee
where $\Xi_i =1 -g^2\, a_i^2$.  The coordinate $t$ is canonically normalised,
and the coordinates $\phi_i$ each have period $2\pi$, in the Kerr-AdS metrics.

   The new metrics that we have obtained, by including the $(n-1)$
parameters $L_\alpha$ in the definition of $X_\alpha$ in (\ref{functions}),
describe the general Kerr-NUT-AdS metrics in dimension $D=2n+1$.  As we shall 
discuss in section \ref{oddplebsec}, in odd dimensions there is actually a 
redundancy among the $(n-1)$ NUT parameters, with one of them being 
trivial.  Thus the total count of non-trivial parameters in the general
Kerr-NUT-AdS metrics in dimension $D=2n+1$ is $2n-1$, which can be thought of 
$n$ rotation parameters, the mass, and $(n-2)$ NUT charges.

\subsection{The even-dimensional case: $D=2n$}

   In this case we begin by defining functions as follows:
\bea
U&=&  \prod_{\alpha=1}^{n-1} (r^2+ y_\alpha^2)\,,\qquad
U_\alpha= - (r^2+ y_\alpha^2){{\prod}'}_{\beta=1}^{\, n-1}
(y_\beta^2-y_\alpha^2)\,,\qquad 1\le \alpha\le n-1
\,,\nn\\
W &=&\prod_{\alpha=1}^{n-1} (1-g^2 y_\alpha^2)\,,\qquad
 \gamma_i = \prod_{\alpha=1}^{n-1} (a_i^2 - y_\alpha^2)\,,
\qquad 1\le i\le n-1\,,\nn\\
X &=& (1+g^2 r^2)\, \prod_{k=1}^{n-1} (r^2 + a_k^2) - 2M\, r\,,\nn\\
X_\alpha &=& -(1-g^2 y_\alpha^2)\, \prod_{k=1}^{n-1}
    (a_k^2 - y_\alpha^2) - 2 L_\alpha\, y_\alpha\,,
\qquad 1\le \alpha\le n-1\,.\label{functionseven}
\eea
We find that the Kerr-NUT-AdS metrics in $D=2n$ dimensions are given by
\bea
ds^2 &=&  \fft{U}{X}\, dr^2
+\sum_{\alpha=1}^{n-1} \fft{U_\alpha}{X_\alpha}\, dy_\alpha^2
- \fft{X}{U}\,
\Big[Wd \td t - \sum_{i=1}^{n-1}  \gamma_i d\td\phi_i\Big]^2\nn\\
&&
+ \sum_{\alpha=1}^{n-1} \fft{X_\alpha}{U_\alpha}\,
\Big[ \fft{(1+g^2 r^2) W}{1- g^2 y_\alpha^2}\, d\td t -
  \sum_{i=1}^{n-1} \fft{(r^2 +a_i^2)\, \gamma_i}{a_i^2- y_\alpha^2}\,
  d\td\phi_i \Big]^2 \,.\label{metriceven}
\eea
Again, the previously-known Kerr-AdS metrics correspond to setting the new
NUT parameters $L_\alpha$ to zero in the definition of the functions
$X_\alpha$ in (\ref{functionseven}).  The coordinates $\td t$ and $\td\phi_i$ 
are related to the canonically-normalised coordinates $t$ and $\phi_i$ of
the $L_\alpha=0$ Kerr-AdS metrics by
\be
t= \td t\, \prod_{i=1}^n \Xi_i\,,\qquad \phi_i= a_i\, \Xi_i\, \td\phi_i\,
       {{\prod}'}_{k=1}^{\, n-1} (a_i^2-a_k^2)\,,\label{ttp2}
\ee
When $L_\alpha=0$, regularity of the Kerr-AdS metric dictates that the 
azimuthal angles $\phi_i$ should all have period $2\pi$.

    As we shall discuss in section 
\ref{evenplebsec}, all the NUT parameters are non-trivial in even dimensions,
and so the general Kerr-NUT-AdS metrics in dimension $D=2n$ have $2n-1$
independent parameters, comprising $(n-1)$ rotations, the mass, and $(n-1)$ 
NUT parameters.

\section{A Simpler Form for the Kerr-NUT-AdS Metrics}\label{simpsec}

   We already saw in section 2 that the Kerr-NUT-AdS metrics assume a
rather symmetrical form when the latitude coordinates $\mu_i$ are
parameterised in terms of the coordinates $y_\alpha$ using (\ref{muy}).
The parallel between the radial coordinate $r$ and the latitude coordinates
$y_\alpha$ becomes even more striking if we perform a Wick rotation of the
radial variable, and define the $n$ coordinates $x_\mu$ by
\be
x_n= \im\, r\,,\qquad x_\alpha= y_\alpha\,,\qquad 1\le\alpha\le n-1\,.
\label{xyr}
\ee
As we shall show, the Kerr-NUT-AdS metrics can now be written in 
a considerably simpler form.  In fact, if we then perform further
transformations on the time and azimuthal coordinates, we arrive at
an even simpler way of presenting the Kerr-NUT-AdS metrics, which 
generalises the four-dimensional expressions obtained in \cite{pleb}.
As always, it is convenient to separate the discussion at this stage 
into the cases of odd and even dimensions. 

\subsection{$D=2n+1$ dimensions}\label{oddplebsec}

   We first define the functions
\bea
U_\mu&=& {{\prod}'}_{\nu=1}^n (x_\nu^2 - x_\mu^2)\,,\qquad
X_\mu = \fft{(1-g^2 x_\mu^2)}{x_\mu^2}\,\prod_{k=1}^n (a_k^2 - x_\mu^2)
          + 2M_\mu\,,\nn\\
\wtd W &=& \prod_{\nu=1}^n (1-g^2 x_\nu^2)\,,\qquad
\td\gamma_i= \prod_{\nu=1}^n(a_i^2-x_\nu^2)\,.\label{2n1X}
\eea
The odd-dimensional Kerr-NUT-AdS metric (\ref{oddmetric}) can then be 
written as
\bea
ds^2 &=&
\sum_{\mu=1}^{n}\Big\{ \fft{U_\mu}{X_\mu}\, dx_\mu^2
+  \fft{X_\mu}{U_\mu}\,
\Big[ \fft{\wtd W}{1- g^2 x_\mu^2}\, d\td t -
  \sum_{i=1}^n \fft{a_i^2 \, \td\gamma_i}{a_i^2- x_\mu^2}\,
  d\td\phi_i \Big]^2\Big\} \nn\\
&&
 - \fft{\prod_{k=1}^n a_k^2}{\prod_{\mu=1}^{n} x_\mu^2}\,
\Big[\wtd W\, d\td t - \sum_{i=1}^n \td \gamma_i\, d\td\phi_i
    \Big]^2\,.\label{oddmetric2}
\eea
Note that $M_n$ is just equal to the previous mass parameter $M$,
while the remaining $M_\alpha$ are NUT parameters, previously
denoted by $L_\alpha$.

   It is useful to give also the inverse of the metric (\ref{oddmetric2}).
Defining
\be
S_\mu = \prod_{k=1}^n (a_k^2-x_\mu^2)^2\,,\qquad B_j = {{\prod}'}_{k=1}^n 
        (a_j^2-a_k^2)\,,
\ee
we find that the inverse metric is given by
\bea
\Big(\fft{\del}{\del s}\Big)^2 &=& \sum_{\mu=1}^n\Big\{
  \fft{X_\mu}{U_\mu}\, \Big(\fft{\del}{\del x_\mu}\Big)^2 +
 \fft{S_\mu}{x_\mu^4\, U_\mu\, X_\mu}\Big[
\fft{1}{(\prod_j \Xi_j)}\, \fft{\del}{\del \td t} +
 \sum_{k=1}^n \fft{(1-g^2 x_\mu^2)}{B_k\, \Xi_k\, 
             (a_k^2 -x_\mu^2)}\, \fft{\del}{\del \td\phi_k}\Big]^2\Big\}\nn\\
&& - \fft{(\prod_{k=1}^n a_k^2)}{(\prod_{\nu=1}^n x_\nu^2)}\,
\Big( \fft1{(\prod_j \Xi_j)}\, \fft{\del}{\del\td t} + 
 \sum_{k=1}^n \fft1{a_k^2 B_k\, \Xi_k}\, \fft{\del}{\del\td\phi_k}
       \Big)^2\,.\label{oddinv1}
\eea
The inverse metric becomes somewhat simpler if expressed in terms of the
original canonically normalised coordinates $t$ and $\phi_k$, whose
relation to $\td t$ and $\td \phi_k$ is given in (\ref{ttp}).  The metric
(\ref{oddinv1}) then becomes 
\bea
\Big(\fft{\del}{\del s}\Big)^2 &=& \sum_{\mu=1}^n\Big\{
  \fft{X_\mu}{U_\mu}\, \Big(\fft{\del}{\del x_\mu}\Big)^2 +
 \fft{S_\mu}{x_\mu^4\, U_\mu\, X_\mu}\Big[
\fft{\del}{\del t} +
 \sum_{k=1}^n \fft{a_k\, (1-g^2 x_\mu^2)}{
             (a_k^2 -x_\mu^2)}\, \fft{\del}{\del \phi_k}\Big]^2\Big\}\nn\\
&& - \fft{(\prod_{k=1}^n a_k^2)}{(\prod_{\nu=1}^n x_\nu^2)}\,
\Big( \fft{\del}{\del t} +
 \sum_{k=1}^n \fft1{a_k }\, \fft{\del}{\del\phi_k}
       \Big)^2\,.\label{oddinv2}
\eea

   It is straightforward to see that the Kerr-NUT-AdS metrics  
(\ref{oddmetric}) and (\ref{oddmetric2}) have a set of discrete 
symmetries under which one of the rotation parameters $a_i$ is
inverted through the AdS radius $1/g$.  Thus, choosing $a_1$ for
this purpose as a representative example, we can see that 
(\ref{oddmetric2}) is invariant under the set of 
transformations
\bea
&&a_1\, g\rightarrow \fft1{a_1\, g}\,,\qquad
 a_j\rightarrow \fft{a_j}{a_1\, g}\,,\quad 2\le j\le n\,,\nn\\
&& M_\mu\rightarrow \fft{M_\mu}{(a_1\, g)^{2n}}\,,\qquad
gt\rightarrow \phi_1\,,\qquad \phi_1 \rightarrow gt\,,
\qquad x_\mu \rightarrow \fft{x_\mu}{a_1\,g}\,,
\eea
with $\phi_j$ for $2\le j\le n$ left unchanged.  This, and the other
permutation-related inversion symmetries, can always map a metric
with over-rotation (one or more parameters $a_i$ satisfying $|a_i\, g|>1$)
into a metric with under-rotation (all parameters satisfying $|a_i\, g|<1$).

    A further simplification of the new Kerr-NUT-AdS metrics 
that we obtained in (\ref{oddmetric}) and (\ref{oddmetric2}) in 
dimensions $D=2n+1$ is possible, allowing them to be written in a 
manner that is a rather natural higher-dimensional analogue of the 
expression in \cite{pleb} for the four-dimensional rotating black
hole metrics.  Again we begin by performing the  ``Wick rotation''
of the radial variable, as in (\ref{xyr}).  We then find that after
appropriate linear redefinitions of the time and azimuthal coordinates, 
the $D=2n+1$ Kerr-NUT-AdS metrics can be written as
\be
ds^2= \sum_{\mu=1}^n \Big\{
         \fft{dx_\mu^2}{Q_\mu} + Q_\mu\, \Big( \sum_{k=0}^{n-1}
          A_\mu^{(k)}\, d\psi_k\Big)^2\Big\} -
           \fft{c}{(\prod_{\nu=1}^n x_\nu^2)}\, 
           \Big( \sum_{k=0}^n A^{(k)} \, d\psi_k\Big)^2
\,,\label{2n1pleb}
\ee
where we define
\bea
Q_\mu &=& \fft{X_\mu}{U_\mu}\,,\qquad U_\mu = {{\prod}'}_{\nu=1}^n
     (x_\nu^2 - x_\mu^2)\,, \qquad X_\mu = \sum_{k=1}^n c_k\, x_\mu^{2k}
         + \fft{c}{x_\mu^2}  - 2 b_\mu\,,\nn\\
A_\mu^{(k)} &=& \sum'_{\nu_1 <\nu_2 <\cdots < \nu_k} 
            x_{\nu_1}^2 x_{\nu_2}^2\cdots
        x_{\nu_k}^2\,,\qquad A^{(k)} = \sum_{\nu_1<\nu_2\cdots <\nu_k}
               x_{\nu_1}^2 x_{\nu_2}^2 \cdots x_{\nu_k}^2\,.\label{QY}
\eea
Here, the prime on the summation symbol in the definition of $A_\mu^k$
indicates that the index value $\mu$ is omitted in the summations of
the $\nu$ indices over the range $[1,n]$. Note that $\psi_0$
plays the r\^ole of the time coordinate.   It is worth remarking that
$A^{(k)}$ and
$A_\mu^{(k)}$ can be defined via the generating functions
\be
\prod_{\nu=1}^{n} (1+ \lambda\, x_\nu^2) = \sum_{k=0}^n \lambda^k\, 
    A^{(k)}\,,\qquad
(1 + \lambda x_\mu^2)^{-1}\, \prod_{\nu=1}^{n} (1+ \lambda\, x_\nu^2) 
= \sum_{k=0}^{n-1} \lambda^k\,A_\mu^{(k)}\,.
\ee

   The constants $c_k$, $c$ and $b_\mu$
are arbitrary, with $c_n= (-1)^{n+1}\, g^2$
determining the value of the cosmological
constant, $R_{\mu\nu} = -2n g^2 g_{\mu\nu}$.  The remaining $2n$ constants
$c_k$, $c$ and $b_\mu$ are related to the $n$ rotation parameters $a_i$, the
mass $M$ and the $(n-1)$ NUT
parameters $L_\alpha$ in the obvious way that follows by comparing $X_\mu$ in 
(\ref{QY}) with $X_\mu$ in (\ref{2n1X}).  However, it should be noted that
not all the parameters are non-trivial in the general solution.  This
can be seen from the fact that there is a scaling symmetry of the 
metric (\ref{2n1pleb}), under which we send
\bea
&& x_\mu\rightarrow \lambda \, x_\mu\,,\qquad \psi_k\rightarrow 
            \lambda^{-2k-1}\, \psi_k\,,\nn\\
&&
c_k\rightarrow \lambda^{2n-2k}\, c_k\,,\qquad 
   c\rightarrow \lambda^{2n+2}\, c\,,\qquad
        b_\mu\rightarrow \lambda^{2n}\, b_\mu\,.\label{oddscale}
\eea
This scaling symmetry implies that there is one trivial parameter in the
general Kerr-NUT-AdS solution, leaving a total of $2n-1$ non-trivial 
parameters in $D=2n+1$ dimensions.  In fact, in odd dimensions there
is not necessarily a clear distinction between rotation parameters and 
NUT parameters, as can be seen by comparing the expressions for the functions
$X_\mu$ in (\ref{QY}), and the expressions in
terms of rotations, mass and NUT parameters in (\ref{2n1X}). 
One might for example find that for
some values of the parameters, if the scaling symmetry (\ref{oddscale}) is
used in order to remove a ``redundant'' NUT charge, then this leads to
a rotation parameter becoming imaginary.  In such a range of the parameters,
it would be more natural to retain the redundant NUT parameter.  This is quite
different from the situation in even dimensions, where the mass and NUT 
parameters are distinguished by being the coefficients of linear powers
of the coordinates, as can be seen in (\ref{functionseven}) and in (\ref{2nX})
below.

   We find that the inverse of the metric (\ref{2n1pleb}) is given by
\bea
\Big(\fft{\del}{\del s}\Big)^2 &=& \sum_{\mu=1}^n\Big\{
Q_\mu\, \Big(\fft{\del}{\del x_\mu}\Big)^2
   + \fft1{x_\mu^4\, Q_\mu\, U_\mu^2}\,
\Big[\sum_{k=0}^{n} (-1)^k\, x_\mu^{2(n-k)}\,
\fft{\del}{\del \psi_k}\Big]^2\Big\}\nn\\
&& -
   \fft1{c\, (\prod_{\nu=1}^n x_\nu^2)}\, 
          \Big(\fft{\del}{\del \psi_n}\Big)^2\,.
\eea

    The specific case of the Kerr-NUT-AdS metric in $D=7$ dimensions 
is discussed in section \ref{d7subsec}, including the explicit
transformation of the time and azimuthal coordinates that brings the
metric into the form (\ref{2n1pleb}).  We also give a more extensive 
discussion of the counting of non-trivial parameters in this example.

\subsection{$D=2n$ dimensions}\label{evenplebsec}

   In this case, in addition to performing the Wick rotation of the radial 
variable as in (\ref{xyr}), one must additionally rescale the mass 
by a factor of $\im$ in order to obtain a real metric.
We then define functions
\bea
U_\mu &=&  {{\prod}'}_{\nu=1}^n (x_\nu^2-x_\mu^2)\,,\qquad
X_\mu = -(1-g^2 x_\mu^2)\, \prod_{k=1}^{n-1}(a_k^2 - x_\mu^2)
   -2M_\mu\, x_\mu
\,,\nn\\
\wtd W &=& \prod_{\nu=1}^n (1-g^2 x_\nu^2)\,,\qquad
\td\gamma_i= \prod_{\nu=1}^n(a_i^2-x_\nu^2)\,,\label{2nX}
\eea
where $M_n= \im\, M$ and $M_\alpha=L_\alpha$.
The even-dimensional Kerr-NUT-AdS metrics (\ref{metriceven}) can then
be written as
\bea
ds^2 &=&
\sum_{\mu=1}^{n}\Big\{ \fft{U_\mu}{X_\mu}\, dx_\mu^2
+ \fft{X_\mu}{U_\mu}\,
\Big[ \fft{\wtd W}{1- g^2 x_\mu^2}\, d\td t -
  \sum_{i=1}^{n-1} \fft{\td\gamma_i}{a_i^2- x_\mu^2}\,
  d\td\phi_i \Big]^2 \Big\}\,.\label{metriceven2}
\eea

   We find that the inverse of the metric (\ref{metriceven2}) is given by
\be
\Big(\fft{\del}{\del s}\Big)^2 =
\sum_{\mu=1}^n\Big\{ 
  \fft{X_\mu}{U_\mu}\, \Big(\fft{\del}{\del x_\mu}\Big)^2 +
\fft{S_\mu}{U_\mu X_\mu}\, 
  \Big[ \fft{1}{(\prod_k \Xi_k)}\, \fft{\del}{\del \td t} +
\sum_{k=1}^{n-1} \fft{1-g^2 x_\mu^2}{\Xi_k \,B_k\, 
(a_k^2-x_\mu^2)}\, \fft{\del}{\del\td\phi_k}\Big]^2 \Big\}\,,
\label{inverseeven1}
\ee
where
\be
S_\mu = \prod_{k=1}^{n-1} (a_k^2 - x_\mu^2)^2\,,\qquad
B_j ={{\prod}'}_{k=1}^{n-1} (a_j^2-a_k^2)
\,.
\ee
Note that in terms of the original canonically-normalised coordinates
$t$ and $\phi_i$, the inverse metric (\ref{inverseeven1}) takes the 
slightly simpler form
\be
\Big(\fft{\del}{\del s}\Big)^2 =
\sum_{\mu=1}^n\Big\{
  \fft{X_\mu}{U_\mu}\, \Big(\fft{\del}{\del x_\mu}\Big)^2 +
\fft{S_\mu}{U_\mu X_\mu}\,
  \Big[ \fft{\del}{\del t} +
\sum_{k=1}^{n-1} \fft{a_k(\, 1-g^2 x_\mu^2)}{
(a_k^2-x_\mu^2)}\, \fft{\del}{\del\phi_k}\Big]^2 \Big\}\,,
\label{inverseeven2}
\ee

      The even-dimensional Kerr-NUT-AdS metrics
(\ref{metriceven}) and (\ref{metriceven2}) also have a set of discrete
symmetries under which any one of the rotation parameters $a_i$ is
inverted through the AdS radius $1/g$.  Thus, for example, 
(\ref{metriceven2}) is invariant under the set of
transformations
\bea
&&a_1\, g\rightarrow \fft1{a_1\, g}\,,\qquad
 a_j\rightarrow \fft{a_j}{a_1\, g}\,,\quad 2\le j\le n-1\,,\nn\\
&& M_\mu\rightarrow \fft{M_\mu}{(a_1\, g)^{2n-1}}\,,\qquad
gt\rightarrow \phi_1\,,\qquad \phi_1 \rightarrow gt\,,
\qquad x_\mu\rightarrow \fft{x_\mu}{a_1\,g}\,,\label{inveven}
\eea
with $\phi_j$ for $2\le j\le n-1$ left unchanged.  This, and the other
permutation-related inversion symmetries, can always map a metric
with over-rotation (one or more parameters $a_i$ satisfying $|a_i\, g|>1$)
into a metric with under-rotation (all parameters satisfying $|a_i\, g|<1$).

   Again, we find that the new Kerr-NUT-AdS metrics in dimension $D=2n$, which
we have obtained in (\ref{metriceven}) and (\ref{metriceven2}), 
can be further simplified and
written elegantly in a
form that is a natural higher-dimensional analogue of the four-dimensional
metrics in \cite{pleb}.  After making the Wick rotation
of the radial variable, as in (\ref{xyr}), we then find that after
appropriate linear redefinitions of the time and azimuthal coordinates, 
the $D=2n$
Kerr-NUT-AdS metrics can be written as
\be
ds^2= \sum_{\mu=1}^n \Big\{ 
         \fft{dx_\mu^2}{Q_\mu} + Q_\mu\, \Big( \sum_{k=0}^{n-1}
          A_\mu^{(k)}\, d\psi_k\Big)^2\Big\}\,,\label{2npleb}
\ee
where we define
\bea
Q_\mu &=& \fft{X_\mu}{U_\mu}\,,\qquad U_\mu = {{\prod}'}_{\nu=1}^n
     (x_\nu^2 - x_\mu^2)\,, \qquad X_\mu = \sum_{k=0}^n c_k\, x_\mu^{2k} +
      2 b_\mu\, x_\mu\,,\nn\\
A_\mu^{(k)} &=& \sum'_{\nu_1 <\nu_2 <\cdots < \nu_k} 
  x_{\nu_1}^2 x_{\nu_2}^2\cdots 
        x_{\nu_k}^2\,.\label{QYeven}
\eea
Again, the prime on the summation symbol in the definition of $A_\mu^k$ 
indicates that the index value $\mu$ is omitted in the summations of
the $\nu$ indices over the range $[1,n]$.  The constants $c_k$ and $b_\mu$
are arbitrary, with $c_n= (-1)^{n+1}\, g^2$ 
determining the value of the cosmological
constant, $R_{\mu\nu} = -(2n-1) g^2 g_{\mu\nu}$.  The remaining constants
$c_k$ and $b_\mu$ are related to the rotation parameters, mass and NUT
parameters in the obvious way that follows by comparing $X_\mu$ in 
(\ref{QYeven}) with $X_\mu$ in (\ref{2nX}).  

    In this even-dimensional 
case there is ostensibly a mismatch between the total number of parameters
in the metrics (\ref{metriceven}) or (\ref{metriceven2}), namely $(n-1)$ 
rotation parameters $a_i$, the mass $M$ and the $(n-1)$ NUT parameters
$L_\alpha$, and the number of parameters in the polynomials $X_\mu$, namely
$n$ constants $c_k$ for $0\le k\le n-1$, and $n$ constants $b_\mu$.  However,
there is also a scaling symmetry that leaves the metric (\ref{2npleb}) 
invariant, namely
\bea
&& x_\mu\rightarrow \lambda \, x_\mu\,,\qquad \psi_k\rightarrow
            \lambda^{-2k-1}\, \psi_k\,,\nn\\
&&
c_k\rightarrow \lambda^{2n-2k}\, c_k\,,\qquad
        b_\mu\rightarrow \lambda^{2n}\, b_\mu\,.
\eea
This implies one parameter in $X_\mu$ is trivial, leaving $2n-1$ non-trivial
parameters in total in the general Kerr-NUT-AdS solution in dimension
$D=2n$.\footnote{It should be emphasised that there is a significant
difference therefore between even and odd dimensions, as regards the 
number of non-trivial NUT charges that can be introduced.  In even dimensions
$D=2n$ the general Kerr-AdS metrics can be augmented with the introduction
of $(n-1)$ non-trivial NUT parameters, while in odd dimensions $D=2n+1$ 
the general Kerr-AdS metrics can be augmented with the introduction of 
$(n-2)$ non-trivial NUT parameters.  Thus, in particular, there is a 
non-trivial NUT charge in $D=4$, but there is no non-trivial NUT charge in
$D=5$.  In odd dimensions, only in $D=7$ and above does one have 
non-trivial NUT charges.}

    It is useful also to record the inverse of the metric (\ref{2npleb}),
which we find to be
\be
\Big(\fft{\del}{\del s}\Big)^2 = \sum_{\mu=1}^n\Big\{
Q_\mu\, \Big(\fft{\del}{\del x_\mu}\Big)^2  
   + \fft1{Q_\mu\, U_\mu^2}\,  
\Big[\sum_{k=0}^{n-1} (-1)^k\, x_\mu^{2(n-1-k)}\, 
\fft{\del}{\del \psi_k}\Big]^2\Big\}\,.
\ee

   The specific case of the Kerr-NUT-AdS metric in $D=6$ dimensions
is discussed in section \ref{d6subsec},  including the explicit transformation
of the time and azimuthal coordinates that brings the metric into the form
(\ref{2npleb}).

\section{Kerr-NUT-AdS Metrics in $D=6$ and $D=7$}

\subsection{Seven-dimensional Kerr-NUT-AdS}\label{d7subsec}

         Here we present the specific example of $D=7$, with rotation
parameters $a_i=\{a, b, c\}$, mass $M$ and two NUT parameters
$L_1$ and $L_2$.  The Kerr-NUT-AdS metric is given by
\bea
ds^2&=& \fft{(r^2 + y^2) (r^2 + z^2)\, dr^2}{X} +
        \fft{(r^2 + y^2) (y^2-z^2)\, dy^2}{Y} +
        \fft{(r^2 + z^2) (z^2-y^2)\, dz^2}{Z} \nn\\
&&-\fft{X}{(r^2+y^2)(r^2+z^2)}\Big[(1-g^2 y^2)(1-g^2 z^2)\,d\td t -
a^2 (a^2-y^2)(a^2-z^2)\,d\td\phi_1\nn\\
&&\qquad\qquad\qquad -
b^2 (b^2-y^2)(b^2-z^2)\,d\td\phi_2 -
c^2 (c^2-y^2)(c^2-z^2)d\td\phi_3\Big]^2\nn\\
&&+\fft{Y}{(r^2+y^2)(y^2-z^2)}\Big[
(1+g^2 r^2)(1-g^2 z^2)\, d\td t -
a^2 (a^2 + r^2)(a^2 - z^2)\,d\td\phi_1\nn\\
&&\qquad\qquad\qquad -
b^2 (b^2 + r^2)(b^2-  z^2)\,d\td\phi_2 -
c^2 (c^2 + r^2)(c^2 - z^2)\,d\td\phi_3\Big]^2\nn\\
&&+\fft{Z}{(r^2+z^2)(z^2-y^2)}\Big[(1+g^2 r^2)(1-g^2 y^2)\,d\td t -
a^2 (a^2 + r^2)(a^2 - y^2)\,d\td\phi_1\nn\\
&&\qquad\qquad\qquad -
b^2 (b^2 + r^2)(b^2-  y^2)\, d\td\phi_2 -
c^2 (c^2 + r^2)(c^2 - y^2)\,d\td\phi_3 \Big]^2\nn\\
&& +\fft{a^2 b^2 c^2}{r^2 y^2 z^2}\Big[
(1 + g^2 r^2)(1-g^2 y^2)(1-g^2 z^2)\,d\td t -
(a^2+r^2)(a^2-y^2)(a^2-z^2)\, d\td\phi_1\nn\\
&&\qquad -
(b^2+r^2)(b^2-y^2)(b^2-z^2)\,d\td\phi_2
 - (c^2+r^2)(c^2-y^2)(c^2-z^2)\, d\td\phi_3\Big]^2\,,
\eea
where
\bea
&&\td t=\fft{t}{\Xi_a\Xi_b\Xi_c}\,,\qquad
\td\phi_1=\fft{\phi_1}{a\,\Xi_a (b^2-a^2)(c^2-a^2)}\,,\nn\\
&&\td\phi_2=\fft{\phi_2}{b\, \Xi_b (a^2-b^2)(c^2-b^2)}\,,\qquad
\td\phi_3=\fft{\phi_3}{c\, \Xi_c (a^2-c^2)(b^2-c^2)}\,,\nn\\
&&\Xi_a=1-a^2 g^2\,,\qquad
\Xi_b=1- b^2 g^2\,,\qquad
\Xi_c=1- c^2 g^2\,,\nn\\
&&
X=\fft{1}{r^2}(1+g^2 r^2)(a^2+r^2)(b^2+r^2)(c^2+r^2) - 2M\,,\nn\\
&&
Y=\fft{1}{y^2}(1-g^2 y^2)(a^2-y^2)(b^2-y^2)(c^2-y^2) + 2L_1\,,\nn\\
&&
Z=\fft{1}{z^2}(1-g^2 z^2)(a^2-z^2)(b^2-z^2)(c^2-z^2) + 2L_2\,.
\eea
Note that regularity of the metric dictates that the coordinates 
$\phi_i$ each have period $2\pi$ when
the NUT parameters $L_1$ and $L_2$ are set to zero.

   The metric has six parameters, $(a, b, c, M, L_1, L_2)$, but
one of them is redundant.  To show this, we first rewrite the metric
after making the coordinate transformations
\bea
t&=&t' + (a^2 + b^2 + c^2) \psi_1 + (a^2b^2 + b^2 c^2 + c^2 a^2)
\psi_2 + a^2 b^2 c^2 \psi_3\,,\nn\\
\fft{\phi_1}{a} &=& \psi_1 + (b^2 + c^2) \psi_2 + b^2 c^2 \psi_3
+ g^2 (t' + (b^2 + c^2) \psi_1 + b^2 c^2 \psi_2)\,,\nn\\
\fft{\phi_2}{b} &=& \psi_1 + (a^2 + c^2) \psi_2 + a^2 c^2 \psi_3
+ g^2 (t' + (a^2 + c^2) \psi_1 + a^2 c^2 \psi_2)\,,\nn\\
\fft{\phi_3}{c} &=& \psi_1 + (a^2 + b^2) \psi_2 + a^2 b^2 \psi_3
+ g^2 (t' + (a^2 + b^2) \psi_1 + a^2 b^2 \psi_2)\,,
\eea
which leads to
\bea
ds^2&=& \fft{(r^2 + y^2) (r^2 + z^2)\, dr^2}{X} +
        \fft{(r^2 + y^2) (y^2-z^2)\, dy^2}{Y} +
        \fft{(r^2 + z^2) (z^2-y^2)\, dz^2}{Z} \nn\\
&&-\fft{X}{(r^2+y^2)(r^2+z^2)}\Big(dt' +
(y^2 + z^2) d\psi_1 + y^2 z^2 d\psi_2\Big)^2\nn\\
&&+\fft{Y}{(r^2+y^2)(z^2-y^2)}\Big(dt' +
(z^2-r^2)d\psi_1 - r^2 z^2 d\psi_2\Big)^2\nn\\
&&+\fft{Z}{(r^2+z^2)(y^2-z^2)}\Big(dt' +
(y^2-r^2)d\psi_1 - r^2 y^2 d\psi_2\Big)^2\\
&& +\fft{C_3}{r^2 y^2 z^2} \Big(dt' +
(y^2 + z^2 - r^2) d\psi_1 + (y^2 z^2 - r^2 y^2 - r^2 z^2)
d\psi_2 - r^2 y^2 z^2 d\psi_3\Big)^2\,.\nn
\eea
The functions $X$, $Y$ and $Z$ can be expressed as
\bea
X &=& g^2 r^6 + C_0 r^4 + C_1 r^2 + C_2-2M + \fft{C_3}{r^2}
\,,\nn\\
Y&=& g^2 y^6 - C_0 y^4 + C_1 y^2 - C_2+2L_1 +\fft{C_3}{y^2}
\,,\nn\\
Z&=& g^2 z^6 - C_0 z^4 + C_1 z^2 - C_2+2L_2 + \fft{C_3}{z^2}
\,,
\eea
where
\bea
C_0&=& 1 + g^2 (a^2 + b^2 + c^2)\,,\qquad
C_1= a^2 + b^2 + c^2 + g^2 (a^2 b^2 + b^2 c^2 + c^2 a^2)\,,\nn\\
C_2&=& a^2b^2 + b^2 c^2 + c^2 a^2 + g^2 a^2b^2 c^2\,,\qquad
C_3= a^2 b^2 c^2\,.
\eea
We can now view the solution as being parameterised by $(C_0, C_1,
C_3)$, together with $X_0=C_2-2M$, $Y_0=2L_1-C_2$, $Z_0=2L_2-C_2$.
The solution has a scaling symmetry, namely
\bea
&&
r \rightarrow \lambda\, r\,,\qquad
y\rightarrow \lambda\, y\,,\qquad
z\rightarrow \lambda\, z\,,\nn\\
&&
C_0\rightarrow \lambda^2 C_0\,,\qquad
C_1\rightarrow \lambda^4 C_1\,,\qquad
C_3\rightarrow \lambda^8 C_3\,,\nn\\
&&
X_0\rightarrow \lambda^6 X_0\,,\qquad
Y_0\rightarrow \lambda^6 Y_0\,,\qquad
Z_0\rightarrow \lambda^6 Z_0\,,\nn\\
&&
\td t\rightarrow \lambda^{-1} \td t\,,\qquad
\psi_1\rightarrow \lambda^{-3} \psi_1\,,\qquad
\psi_2\rightarrow \lambda^{-5} \psi_2\,,\qquad
\psi_3\rightarrow \lambda^{-7} \psi_3\,,\label{paras}
\eea
This implies that one of the parameters in (\ref{paras})
can be set to 1 without
loss of generality. In turn, this allows us to set one of the original
parameters, say $L_2$ to zero.  Thus there are actually five
non-trivial parameters in the solution.

          For any fixed gauged choice of $M, L_1, L_2$, the metric
still has discrete residual symmetry, namely
\be
a\rightarrow \fft{1}{a\, g^2}\,,\qquad
b\rightarrow \fft{b}{a\, g}\,,\qquad
c\rightarrow \fft{c}{a\,g}\,,\qquad
\{M, L_1, L_2\} \rightarrow \lambda^6\, \{M, L_1, L_2\}\,,
\ee
with $\lambda = 1/(a\, g)$.

\subsection{Six-dimensional Kerr-NUT-AdS}\label{d6subsec}

Here we present the explicit $D=6$ metric, given by
\bea
ds^2&=&\fft{(r^2+y^2)(r^2+z^2)\,dr^2}{X} +
\fft{(r^2+y^2)(y^2-z^2)\,dy^2}{Y} +
\fft{(r^2+z^2)(z^2-y^2)\,dz^2}{Z}\nn\\
&&-\fft{X}{(r^2+y^2)(r^2+z^2)} \Big( (1-g^2 y^2)(1-g^2 z^2)\, d\td t -
(a^2-y^2)(a^2-z^2)\, d\td \phi_1\nn\\
&& \qquad\qquad\qquad\qquad\qquad
-(b^2-y^2)(b^2-z^2)\, d\td\phi_2\Big)^2\nn\\
&&+\fft{Y}{(r^2+y^2)(y^2-z^2)} \Big( (1 + g^2 r^2)(1-g^2 z^2)\,  d\td t -
(a^2 + r^2)(a^2-z^2)\, d\td \phi_1\nn\\
&&\qquad\qquad\qquad\qquad\qquad
- (b^2 + r^2) (b^2-z^2)\, d\td \phi_2\Big)^2\nn\\
&&+\fft{Z}{(r^2 + z^2)(z^2-y^2)} \Big( (1 + g^2 r^2)(1-g^2 y^2)\, d\td t -
(a^2 +r^2)(a^2-y^2)\, d\td \phi_1\nn\\
&&\qquad\qquad\qquad\qquad\qquad
- (b^2+r^2)(b^2-y^2)\, d\td \phi_2\Big)^2\,.
\eea
where $X, Y$ and $Z$ are given by
\bea
X\!\!\!&=&\!\!\!(1+ g^2 r^2) (r^2+a^2)(r^2+b^2) - 2M\, r\,,\quad
Y=-(1-g^2y^2)(a^2-y^2)(b^2-y^2) - 2L_1\, y\,,\nn\\
Z\!\!\!&=&\!\!\!-(1-g^2z^2)(a^2-z^2)(b^2-z^2) - 2 L_2\, z\,.\label{d6xyz}
\eea
The coordinate $\td t$, $\td \phi_1$ and $\td \phi_2$ are related to
the canonically defined $t$, $\phi_1$ and $\phi_2$ by (\ref{ttp2}).
We can then make the coordinate transformation
\bea
t&=&t' + (a^2 + b^2)\, \psi_1 + a^2b^2\, \psi_2\,,\qquad
\fft{\phi_1}{a} = \psi_1 + b^2\, \psi_2 + g^2 (d\td t + b^2\, \psi_1)
\,,\nn\\
\fft{\phi_2}{b} &=& \psi_1 + a^2\, \psi_2 + g^2 (d\td t + a^2\, \psi_1)
\,,
\eea
which leads to the metric
\bea
ds^2&=&\fft{(r^2+y^2)(r^2+z^2)\,dr^2}{X} +
\fft{(r^2+y^2)(y^2-z^2)\,dy^2}{Y} +
\fft{(r^2+z^2)(z^2-y^2)\,dz^2}{Z}\nn\\
&&-\fft{X}{(r^2+y^2)(r^2+z^2)}\Big(dt' + (y^2+z^2)\, d\psi_1 +
                                 y^2 z^2\, d\psi_2\Big)^2\nn\\
&&+\fft{Y}{(r^2+y^2)(y^2-z^2)}\Big(dt' + (z^2-r^2)\, d\psi_1 -
                                 r^2 z^2\, d\psi_2\Big)^2\nn\\
&&+\fft{Z}{(r^2+z^2)(z^2-y^2)}\Big(dt' + (y^2-r^2)\, d\psi_1 -
                                 r^2 y^2\, d\psi_2\Big)^2\,.\label{d6pleb}
\eea
The functions $X, Y$ and $Z$ given in (\ref{d6xyz}) can now be written as
\bea
X&=& g^6 r^6 + C_0\, r^4 + C_1 r^2 -2M\, r + C_2\,,\nn\\
Y&=& g^6 y^6 - C_0\, y^4 + C_1 y^2 -2L_1\, y - C_2\,,\nn\\
Z&=& g^6 z^6 - C_0\, z^4 + C_1 z^2 -2L_1\, z - C_2\,,\nn
\eea
where $C_i$ are constants, expressed in terms two constants $a$ and $b$,
given by
\be
C_0=1 + g^2 (a^2 + b^2)\,,\qquad
C_1=a^2 + b^2 + g^2 a^2 b^2\,,\qquad
C_2=a^2 b^2\,.\label{d6cs}
\ee
In fact, the constants $C_i$ can be arbitrary, since the form of the
metric has the following symmetry:
\bea
&&r\rightarrow \lambda\, r\,,\qquad y\rightarrow \lambda\, y\,,\qquad
z\rightarrow \lambda\, z\,,\nn\\
&&
C_0\rightarrow \lambda^2\, C_0\,,\qquad
C_1\rightarrow \lambda^4\, C_1\,,\qquad
C_2\rightarrow \lambda^6\, C_2\,,\nn\\
&&
M\rightarrow \lambda^5\, M\,,\qquad
L_1\rightarrow \lambda^5\, L_1\,,\qquad
L_2\rightarrow \lambda^5\, L_2\,,\nn\\
&&
\td t\rightarrow \lambda^{-1}\, \td t\,,\qquad
\psi_1\rightarrow \lambda^{-3}\, \psi_1\,,\qquad
\psi_2\rightarrow \lambda^{-5}\, \psi_2\,.
\eea
Thus to fix $C_i$ as given by (\ref{d6cs}) is to have fixed the
symmetry.  It follows that unlike in the case of odd dimensions,
the NUT parameters here are all non-trivial.   For the above fixed
parameter gauge, the metric has residual discrete symmetry, namely
\be
a\rightarrow \fft{1}{a\, g^2}\,,\qquad b\rightarrow \fft{b}{a\, g}
\ee
with $\lambda=a\, g$.

   The form in which the six-dimensional Kerr-NUT-AdS metric is written
in equation (\ref{d6pleb}) is closely analogous to the form of the 
four-dimensional Plebanski metrics \cite{pleb}.  

\section{BPS Limits}

\subsection{BPS limit for $D=2n+1$}

   In this section we shall investigate the BPS limit of the odd-dimensional
Kerr-NUT-AdS metrics.  In this limit the metrics admit Killing spinors,
and if one furthermore performs a Euclideanisation to positive-definite
metric signature, and sets the cosmological constant to be positive
(by taking $g^2$ to be negative) one will obtain Einstein-Sasaki metrics.

   For convenience, we shall scale the metrics in this limit so that
their Ricci tensor is
the same as that of a unit sphere of the same dimension.  This is
achieved by setting $g=\im$.  It is convenient also to
write the metrics in a specific asymptotically-rotating frame, by
sending $\phi_i\rightarrow \phi_i -
g\, dt$.

   We shall first consider the 7-dimensional metric discussed in the
previous section.  The Euclideanisation is achieved by sending
\be
t\rightarrow {\rm i}\, \tau\,,\qquad
a\rightarrow {\rm i}\, a\,,\qquad
b\rightarrow {\rm i}\, b\,,\qquad
c\rightarrow {\rm i}\, c\,.
\ee
To take the BPS limit we define
\bea
&&1-a^2 = \alpha\, \epsilon\,,\qquad
1-b^2 = \beta\, \epsilon\,,\qquad
1-c^2 = \gamma\, \epsilon\,,\nn\\
&&
1-r^2 = x\, \epsilon\,,\qquad
1+y^2\rightarrow y\, \epsilon\,,\qquad
1+z^2\rightarrow z\, \epsilon\,,\nn\\
&&
M=m\, \epsilon^4\,,\qquad
L_1=\ell_1\, \epsilon^4\,,\qquad
L_2=\ell_2\, \epsilon^4\,,
\eea
and then send $\epsilon\rightarrow 0$.  This leads to the
metric
\be
ds_7^2 = (d\tau + {\cal A})^2 + ds_6^2\,,
\ee
where
\bea
ds_6^2 &=& \fft{(x-y)(x-z)\, dx^2}{4X} +
\fft{(y-x)(y-z)\, dy^2}{4Y} + \fft{(z-x)(z-y)}{4Z}dz^2\nn\\
&&+\fft{X}{(x-y)(x-z)} \Big((\alpha - y)(\alpha -z) d\td\phi_1 +
(\beta - y)(\beta -z) d\td\phi_2 +
(\gamma - y)(\gamma -z) d\td\phi_3\Big)^2\nn\\
&&+\fft{Y}{(y-x)(y-z)} \Big((\alpha - x)(\alpha -z) d\td\phi_1 +
(\beta - x)(\beta -z) d\td\phi_2 +
(\gamma - x)(\gamma -z) d\td\phi_3\Big)^2\nn\\
&&+\fft{Z}{(z-x)(z-y)} \Big((\alpha - x)(\alpha -y) d\td\phi_1 +
(\beta - x)(\beta -y) d\td\phi_2 +
(\gamma - x)(\gamma -y) d\td\phi_3\Big)^2\nn\\
{\cal A}&=& (\alpha-x)(\alpha-y)(\alpha-z) d\td\phi_1 +
(\beta-x)(\beta-y)(\beta-z) d\td\phi_1 +
(\gamma-x)(\gamma-y)(\gamma-z) d\td\phi_1\nn\\
X&=&x(\alpha -x)(\beta -x)(\gamma -x) - 2m\,,\qquad
Y=y(\alpha -y)(\beta -y)(\gamma -y) - 2\ell_1\,,\nn\\
Z&=&z(\alpha -z)(\beta -z)(\gamma -z) - 2\ell_2\,.\label{d6ek}
\eea
The $\td\phi_i$ are related to the original $\phi_i$ by the
constant scalings
\bea
\phi_1&=&\alpha(\alpha-\beta)(\alpha-\gamma) \td\phi_1\,,\qquad
\phi_2=\beta(\beta-\alpha)(\beta-\gamma)\td\phi_2\,,\nn\\
\phi_3&=&\gamma(\gamma-\alpha)(\gamma-\beta)\td\phi_3\,.
\eea

   For the general case of $D=2n+1$ dimensions, we find after performing
analogous computations that the Einstein-Sasaki metric is given by
\be
ds_{2n+1}^2 = (d\tau + {\cal A})^2 + ds_{2n}^2\,,\label{genes0}
\ee
where
\bea
ds_{2n}^2 &=& \sum_{\mu=1}^n
 \fft{U_\mu\, dx_\mu^2}{4X_\mu} +
\sum_{\mu=1}^n \fft{X_\mu}{U_\mu}
\Big( \sum_{i=1}^n \fft{W_i\, d\td\phi_i}{\alpha_i-x_\mu}
\Big)^2\,,\nn\\
{\cal A} &=& \sum_{i=1}^n W_i d\td\phi_i\,,\qquad
U_\mu={{\prod}'}_{\nu=1}^n
  (x_\nu-x_\mu)\,,\nn\\
X_\mu &=& x_\mu \prod_{i=1}^n (\alpha_i - x_\mu) - 2\ell_\mu\,,\qquad
W_i= \prod_{\nu=1}^n (\alpha_i - x_\nu)\label{genes}
\eea
Thus we obtain a large class of local Einstein-Sasaki metrics in
arbitrary $(2n+1)$ dimensions.  These metrics extend the results
obtained in \cite{clpp1,clpp2}, where there were no NUT charges, 
and those in \cite{lpv,chlupo1}, where metrics of 
cohomogeneity two were considered.  We expect that 
(\ref{genes0},\ref{genes}) is the most
general metric for Einstein-Sasaki spaces with $U(1)^{n+1}$ isometry in
$(2n+1)$ dimensions.

     It is of considerable interest to study the global structure of
these Einstein-Sasaki metrics, and thereby to obtain the conditions on
the parameters under which they extend onto smooth manifolds.
This was done for $D=5$ in \cite{clpp1,clpp2}, where complete metrics
for the Einstein-Sasaki manifolds $L^{pqr}$ were obtained.  Those 
results extended previous results for the $Y^{pq}$ \cite{ypq} 
manifolds, which corresponded to the specialisation where the two 
angular momentum parameters were set equal.   For seven dimensions, 
the global structure has been previously discussed for various special
cases.  When $Y$ and $Z$ in (\ref{d6ek}) both have
a double root, the solution reduces to that obtained in \cite{clpv},
where the global structure was analysed in detail.  If two angular
momenta are set equal, the solution reduces to that obtained
in \cite{lpv} where the global structure was also discussed.  Aside from
these special cases, our general results in $D=7$ that we have obtained 
in this paper are new.  Similarly, our results in $D\ge9$ extend those
obtained previously.

\subsection{BPS limit for $ D = 2 n $}

The BPS limit in this case can give rise to Ricci-flat K\"ahler metrics.
Consider first the example of the six-dimensional Kerr-NUT-AdS metric. 
We perform a Euclideanisation and take an analogous BPS limit to the one we 
discussed above for the seven-dimensional case, by setting
\be M = m \epsilon^3, \qquad L_1 = i \ell_1 \epsilon^3,\qquad L_2 = i
\ell_2 \epsilon^3\,. \ee
In the BPS limit,  when $\epsilon$ goes to zero, we obtain the 
Ricci flat metric
\bea ds_6^2 &=& \fft{(y-x)(z-x)\, dx^2}{4X} +
\fft{(x-y)(z-y)\, dy^2}{4Y} + \fft{(x-z)(y-z)}{4Z}dz^2\nn\\
&&+\fft{X}{(y-x)(z-x)} \Big( y \: z \; d\td\tau - (\alpha -
y)(\alpha
-z) d\td\phi_1 - (\beta - y)(\beta -z) d\td\phi_2 \Big)^2\nn\\
&&+\fft{Y}{(x-y)(z-y)} \Big( x \: z \; d\td\tau - (\alpha -
x)(\alpha
-z) d\td\phi_1 - (\beta - x)(\beta -z) d\td\phi_2 \Big)^2\nn\\
&&+\fft{Z}{(x-z)(y-z)} \Big( x \: y \; d\td\tau - (\alpha -
x)(\alpha
-y) d\td\phi_1 - (\beta - x)(\beta -y) d\td\phi_2 \Big)^2\nn\\
X&=&x(\alpha -x)(\beta -x) - 2m\,,\qquad
Y=y(\alpha -y)(\beta -y) - 2\ell_1\,,\nn\\
Z&=&z(\alpha -z)(\beta -z) - 2\ell_2\,. \eea
The coordinates $\td\tau$ and $\td\phi_i$ are related to the original $\tau$
and $\phi_i$ by the constant scalings
\bea \phi_1&=&\alpha(\alpha-\beta) \td\phi_1\,,\qquad
\phi_2=\beta(\beta-\alpha) \td\phi_2 \,,\qquad \tau = \alpha \beta
\;\td\tau \,. \eea
Note that this metric can in fact be viewed as the zero cosmological 
constant limit of the six-dimensional Einstein-K\"ahler metric (\ref{d6ek})
that we obtained above.

 For the general case of $D=2n$ dimensions, we find that the BPS limit of the
Euclideanised Kerr-NUT-AdS metrics yields the Ricci-flat metrics 
\bea ds_{2n}^2 &=& \sum_{\mu=1}^n  \fft{U_\mu\, dx_\mu^2}{4X_\mu}
   + \sum_{\mu=1}^n
\fft{X_\mu}{U_\mu} \Big(
\fft{\gamma}{x_{\mu}} d\td\tau - \sum_{i=1}^{n-1} \fft{W_i\,
d\td\phi_i}{\alpha_i-x_\mu}
\Big)^2\,,\nn\\
X_\mu &=& x_\mu \prod_{i=1}^{n-1} (\alpha_i - x_\mu) -
2\ell_\mu\,,\qquad U_\mu= {{\prod}'}^n_{\nu=1} (x_\nu-x_\mu)\,,\nn\\
W_i &=& \prod_{\nu=1}^n (\alpha_i -
x_\nu)\,,\qquad \gamma = \prod_{\nu=1}^n x_{\nu}\,.
\eea
Again, these metrics can be obtained also as limiting cases of the
metrics (\ref{genes}), in which the cosmological constant is sent
to zero.

\section{Conclusions}

    In this paper, we have obtained new results for the inclusion of
NUT parameters in the Kerr-AdS metrics that were constructed in
\cite{gilupapo1,gilupapo2}.  Our strategy for doing this involved first
making a judicious choice of coordinates parameterising the latitude
variables in the Kerr-AdS metrics.  By making a change of variables 
analogous to one considered long ago by Jacobi in the theory of
constrained dynamical systems, we were able to rewrite the Kerr-AdS 
solutions of \cite{gilupapo1,gilupapo2} in such a way that the metrics  
become diagonal in a set of unconstrained latitude coordinates $y_\alpha$.
These coordinates then appear in a manner that closely parallels that of the
radial variable $r$, and this immediately suggests a natural generalisation
of the Kerr-AdS metrics to include NUT charges.  We explicitly verified
that the generalised metrics solve the Einstein equations in all dimensions
$D\le 15$, and we expect that they likewise solve the equations in all
higher dimensions.  After further changes of variable, we arrived at
the very simple expressions (\ref{2n1pleb}) and (\ref{2npleb}) for the
general Kerr-NUT-AdS metrics in all odd and even dimensions.  These
expressions can be thought of as natural generalisations of the 
four-dimensional results obtained in \cite{pleb}.

   The general Kerr-NUT-AdS metrics that we have obtained in this 
paper have a total of $(2n-1)$ non-trivial parameters, where the
spacetime dimension is $D=2n+1$ in the odd-dimensional case, and
$D=2n$ in the even-dimensional case.  In odd dimensions these parameters 
can be viewed as comprising $n$ rotations, a mass, and $(n-2)$
NUT charges.  In even dimensions they instead comprise $(n-1)$ rotations,
a mass, and $(n-1)$ NUT charges.  In odd dimensions, but not in even 
dimensions, there is some measure of arbitrariness in the interpretation
of parameters as rotations or NUT charges. 

   An interesting feature of the Kerr-AdS and Kerr-NUT-AdS metrics that
is uncovered by our work is that in all dimensions there exist 
discrete symmetries of the metrics in which one of the rotation parameters
is inverted through the AdS radius $1/g$, together with appropriate scalings
of the other rotation parameters, the mass and the NUT charges.  An 
implication of these symmetries is that any metric with over-rotation, 
\ie where one or more of the rotation parameters exceeds the AdS radius,
is identical, up to coordinate transformations, to a metric with
only under-rotation.  This was observed previously for the Kerr-AdS 
metric in $D=5$ \cite{chlupo1}.  The inversion symmetry was apparently 
not previously noticed in the four-dimensional Kerr-AdS metric, and 
we have presented it explicitly, in the standard coordinate system, in
appendix B.

   We also considered the BPS, or supersymmetric, limits of the Kerr-NUT-AdS
metrics.  In odd dimensions these yield, after Euclideanisation, new 
examples of Einstein-Sasaki metrics.  We expect that by making appropriate
choices for the various parameters in the solutions, one can obtain
new examples of complete Einstein-Sasaki spaces defined on non-singular
compact manifolds.

\section*{Acknowledgement}

We are grateful to Gary Gibbons and Malcolm Perry for useful discussions.
C.N.P. is grateful to the Relativity and Cosmology group at the C.M.S., 
Cambridge, for hospitality during part of this work.

\appendix

\section{A symmetry between the time and azimuthal coordinates}

   It can be observed from the expressions for the Kerr-NUT-AdS metrics
that we obtained in section \ref{simpsec} that the time coordinate and
the azimuthal angular coordinates appear on a very parallel footing.
It is possible, therefore, to present further simplifications of the
expressions (\ref{oddmetric2}) and (\ref{metriceven2}) in odd and
even dimensions that exploit this observation.

    For the
Kerr-NUT-AdS metrics in odd dimensions $D=2n+1$, we make the definitions
\bea a_0 &=& \fft1{g}\,,\qquad \Gamma_I= \prod_{\nu=1}^n (a_I^2 -
x_\nu^2)\,,
\quad 0\le I\le n\,,\nn\\
\td\phi_0 &=& - g^{2n}\, \td t\,,\qquad X_\mu=
\fft{g^2}{x_\mu^2}\, \prod_{I=0}^{n} (a_I^2 -x_\mu^2) + 2M_\mu\,.
\eea
The metric (\ref{oddmetric2}) can then be written as
\be 
ds^2 = \sum_{\mu=1}^{n} \Big\{ \fft{U_\mu}{X_\mu}\, dx_\mu^2 +
\fft{X_\mu}{U_\mu}\, \Big( \sum_{I=0}^{n}
\fft{a_I^2 \, \Gamma_I\, d\td\phi_I}{a_I^2-x_\mu^2} \Big)^2\Big\} 
- \fft{(\prod_{k=1}^n a_k^2)}{(\prod_{\mu=1}^{n} x_\mu^2)}\,
\Big(\sum_{I=0}^{n} \Gamma_I\, d\td\phi_I
    \Big)^2\,.
\ee

    For the
Kerr-NUT-AdS metrics in even dimensions $D=2n$, we make the definitions
\bea
a_0 &=& \fft1{g}\,,\qquad \Gamma_I= \prod_{\nu=1}^n (a_I^2 - x_\nu^2)\,,
\quad 0\le I\le n-1\,,\nn\\
\td\phi_0 &=& - g^{2n-2}\, \td t\,,\qquad
X_\mu= -g^2\, \prod_{I=0}^{n-1} (a_I^2 -x_\mu^2) - 2M_\mu\, x_\mu\,.
\eea
The metric (\ref{metriceven2}) can then be written as
\be
ds^2 = \sum_{\mu=1}^n \Big\{ \fft{U_\mu}{X_\mu}\, dx_\mu^2 +
\fft{X_\mu}{U_\mu}\,   \Big( \sum_{I=0}^{n-1}
\fft{\Gamma_I\, d\td\phi_I}{a_I^2-x_\mu^2}\Big)^2\Big\}\,.
\ee

\section{Inversion symmetry of the $D=4$ rotating black hole}

   We made the observation n section \ref{simpsec} that there exists 
an inversion 
symmetry in all the Kerr-NUT-AdS metrics, in which one of the rotation
parameters is inverted through the AdS radius, together with
corresponding scalings of the other parameters.  A case of particular
interest is in four dimensions.  (The inversion symmetry for the
five-dimensional Kerr-AdS metric was given in \cite{chlupo1}.)
The four-dimensional Kerr-AdS metric can be written as
\bea
ds^2&=&\fft{\rho^2}{\Delta_r} dr^2 + \fft{\rho^2}{\Delta_\theta}\,
d\theta^2-\fft{\Delta_r}{\rho^2} (dt -
\fft{a}{\Xi} \sin^2\theta\, d\phi^2)^2
+\fft{\Delta_\theta\,\sin^2\theta}{\rho^2}(a\, dt -
\fft{r^2+a^2}{\Xi} d\phi)^2\,,\nn\\
\rho^2&=& r^2 + a^2 \cos^2\theta\,,\quad
\Delta_r=(1+g^2 r^2) (r^2 + a^2) - 2 M\, r\,,\quad
\Delta_\theta=1-a^2 g^2 \cos^2\theta\,, \label{d4kerr}
\eea
where $\Xi=1-a^2 g^2$.  It is straightforward to verify that
the metric is invariant under the transformation
\bea
&&a\rightarrow \fft{1}{a\, g^2}\,,\qquad M\rightarrow \fft{M}{a^3 g^3}\,,\nn\\
&&r\rightarrow \fft{r}{a\, g}\,,\quad
\cos\theta\rightarrow ag\, \cos\theta \,,\quad
\phi\rightarrow -\fft{\phi}{a\, g}\,,\quad
t\rightarrow a\,g\, t + \fft{\phi}{g}\,.\label{d4trans}
\eea
Note that the metric (\ref{d4kerr}) is written in a frame that is
asymptotically rotating at infinity.  As a consequence the required 
ignorable coordinate transformations in (\ref{d4trans}) that bring the 
transformed
metric back to its original form do not, unlike those given in 
(\ref{inveven}) for an asymptotically-static frame, simply involve an 
exchange of the azimuthal coordinate and $g$ times the time coordinate. 
If we define an asymptotically-static frame by replacing the azimuthal 
coordinate with $\hat\phi=\phi+ a g^2 t$, then the last two transformations
in (\ref{d4trans}) become simply
\be
\hat\phi\rightarrow g t\,,\qquad g t \rightarrow \hat\phi\,.
\ee


\begin{thebibliography}{99}

\bm{plebdemi} J.F. Plebanski and M. Demianski,
{\it Rotating, charged, and uniformly accelerating mass in general
relativity},
Annals Phys.  {\bf 98} (1976) 98.


\bm{myeper} R.C. Myers and M.J. Perry,
{\it Black holes in higher dimensional space-times},
Annals Phys.  {\bf 172}, 304 (1986).

\bm{hawhuntay} S.W. Hawking, C.J. Hunter and M.M. Taylor-Robinson,
{\it Rotation and the AdS/CFT correspondence},
Phys. Rev. {\bf D59}, 064005 (1999), hep-th/9811056.

\bm{gilupapo1} G.W. Gibbons, H. L\"u, D.N. Page and C.N. Pope,
{\it The general Kerr-de Sitter metrics in all dimensions},
J. Geom. Phys. {\bf 53}, 49 (2005), hep-th/0404008.

\bm{gilupapo2} G.W. Gibbons, H. L\"u, D.N. Page and C.N. Pope,
{\it Rotating black holes in higher dimensions with a cosmological
  constant},
Phys. Rev. Lett.\  {\bf 93}, 171102 (2004), hep-th/0409155.

\bm{chlupo1} W. Chen, H. L\"u and C.N. Pope,
{\it Kerr-de Sitter black holes with NUT charges},
hep-th/0601002.

\bm{mann} R.B. Mann and C. Stelea, {\it New multiply nutty spacetimes,}
  Phys. Lett. B {\bf 634}, 448 (2006) hep-th/0508203.

\bm{jacobi} C. Jacobi, {\it Vorlesungen \"uber Dynamik}, Reimer, Berlin
(1866).

\bm{pleb} J.F. Plebanski, {\it A class of solutions of
Einstein-Maxwell equations}, Ann. Phys. {\bf 90}, 196 (1975).

\bm{clpp1}
M. Cveti\v c, H.L\"u, D.N. Page and C.N. Pope,
{\it New Einstein-Sasaki spaces in five and higher dimensions,}
  Phys. Rev. Lett.  {\bf 95}, 071101 (2005), hep-th/0504225.

\bm{clpp2}
M. Cveti\v c, H. L\"u, D.N. Page and C.N. Pope,
{\it New Einstein-Sasaki and Einstein spaces from Kerr-de Sitter,}
hep-th/0505223.

\bm{lpv}
H. L\"u, C.N. Pope and J.F. V\'azquez-Poritz,
{\it A new construction of Einstein-Sasaki metrics in $D\ge 7$},
hep-th/0512306.

\bibitem{ypq} J.P. Gauntlett, D. Martelli, J. Sparks and D. Waldram,
{\it Sasaki-Einstein metrics on $S^2 \times S^3$}
Adv. Theor. Math. Phys.\  {\bf 8}, 711 (2004), hep-th/0403002.


\bibitem{clpv} W. Chen, H. L\"u, C.N. Pope and J.F. V\'azquez-Poritz,
{\it A note on Einstein-Sasaki metrics in $D \ge 7$},
  Class. Quant. Grav.  {\bf 22}, 3421 (2005), hep-th/0411218.

\end{thebibliography}
\end{document}